\documentclass{aa}  
\usepackage{graphicx}
\usepackage{txfonts}

\def\bc{{{BRITE}-Constellation}}
\def\b{{\sc {BRITE}}}
\def\bab{BAb}
\def\blb{BLb}
\def\btr{BTr}

\def\deg{$^{\circ}$}

\def\prot{$P_{\rm rot}$}
\def\ac{$\alpha$\,Cir}

\def\Teff{$T_{\rm{eff}}$}
\def\lg{log\,$g$}
\def\hd{HD\,128898}

\def\cd{\rm d$^{-1}$}

\def\w{{\sc {WIRE}}}
\def\s{{\sc {SMEI}}}
\def\t{{\sc {TESS}}}

\def\f1{$f_{1}$}
\def\f7{$f_{7}$}

\def\cms{[cm\,s$^{-2}$]}

\begin{document} 

   \title{New BRITE-Constellation\thanks{Based on data collected by the BRITE-Constellation satellite mission, designed, built, launched, operated and supported by the Austrian Research Promotion Agency (FFG), the University of Vienna, the Technical University of Graz, the University of Innsbruck, the Canadian Space Agency (CSA), the University of Toronto Institute for Aerospace Studies (UTIAS), the Foundation for Polish Science \& Technology (FNiTP MNiSW), and National Science Centre (NCN). } observations of the roAp star \ac
        }


\author {W.\,W. Weiss \inst{1}
\and H.-E. Fr\"{o}hlich\inst{2}
\and T. Kallinger\inst{1}
\and R. Kuschnig\inst{1, 3}
\and A. Popowicz \inst{4}     
\and D. Baade\inst{5}
\and D. Buzasi \inst{6} 
\and G. Handler\inst{7}
\and \makebox{O. Kochukhov\inst{8}}
\and O. Koudelka\inst{3}         
\and A.\,F.\,J.\,Moffat\inst{9}
\and B. Pablo\inst{10}
\and G. Wade\inst{11}        
\and K. Zwintz\inst{12}
}
                        
\institute{University of Vienna, Institute for Astrophysics, T\"urkenschanzstrasse 17, A-1180 Vienna, Austria; \\\email{werner.weiss@univie.ac.at} 
\and Kleine Strasse 9, D-14482 Potsdam;  
\and Graz University of Technology, Institute of Commun. Networks and Satellite Commun., Infeldgasse 12, 8010 Graz, Austria;  %
\and Silesian University of Technology, Department of Electronics, Electrical Engineering and Microelectronics, Poland;  
\and ESO, 
Karl-Schwarzschild-Str. 2, 85748 Garching b. M\"unchen, Germany; 
\and Florida Gulf Coast University, Department of Chemistry and Physics, 
10501 FGCU Blvd S, Fort Myers, FL 33965 USA; 
\and Nicolaus Copernicus Astronomical Center of the Polish Academy of Sciences, Bartycka 18, 00-716 Warsaw, Poland;
\and Uppsala University, Department of Physics and Astronomy, 
Box 516, 75120 Uppsala, Sweden; 
\and Universit\'e de Montr\'eal, Dept. de physique, CP 6128 Succ. C-V, Montr\'eal, QC H3C 3J7, Canada;    
\and American Association of Variable Star Observers, 49 Bay State Rd., Cambridge MA 02138, USA; 
\and Royal Military College, Dept. of Physics and Space Science, PO Box 17000 Station Forces, Kingston, ON, K7K 0C6 Canada;
\and University of Innsbruck, Institute for Astro- and Particle Physics, Technikerstrasse 25, A-6020 Innsbruck   
}

\date{Received ; accepted }

\abstract
        {Chemically peculiar (CP) stars with a measurable magnetic field comprise the group of mCP stars. The pulsating members define the subgroup of rapidly oscillating Ap (roAp) stars, of which \ac\ is the brightest member.  Hence, \ac\ allows the application of challenging techniques, such as  interferometry, very high temporal and spectral resolution photometry, and spectroscopy in a wide wavelength range, that have the potential to provide unique information about the structure and evolution of a star.}
        {Based on new photometry from \bc, obtained with blue and red filters, and on photometry from \w , \s , and \t\ we attempt to determine the surface spot structure of \ac\  and investigate pulsation frequencies.}
        {We used photometric surface imaging and frequency analyses and  Bayesian techniques in order to  quantitatively compare the probability of different models. }
        {\bc\  photometry obtained from 2014 to 2016 is put in the context of space photometry obtained by  \w , \s , and \t . This provides  improvements in the determination of the   rotation period and  surface features (three spots detected and a fourth one indicated).  The main pulsation frequencies indicate two consecutive radial modes and one intermediate dipolar mode.  Advantages and problems of the applied Bayesian technique are discussed.}
        {}
        

   \maketitle

\section{Introduction}

Rapidly oscillating chemically peculiar (roAp) stars are a subgroup of the chemically peculiar (CP) stars with  spectral types ranging from late B to early F, with luminosity class IV to V, and a global magnetic field that causes a peculiar chemical composition of their atmosphere and that deviates from   solar composition \citep{kurtz3}. The roAp stars provide important tools for investigating stellar structure and evolution and for testing   astrophysical concepts. These stars show low-degree, high-order acoustic oscillations, similar to the 5 min oscillations in the Sun, but with coupling to the magnetic field. The excitation of such high-order oscillations, instead of the low-order pulsation in $\delta$\,Scuti stars, is expected to be related to the magnetic field, which suppresses the near-surface convection, and therefore increases the efficiency of the opacity mechanism in the hydrogen ionisation region \citep[][]{ balmforth2001}. 
In principle, the high-order oscillations should simplify the theoretical interpretation of the observed power spectrum by the use of standard asymptotic tools, but since the oscillations are modified near the surface by the magnetic field,  more sophisticated modelling is required \citep[][]{cun2006}. However, the coupling between rotation, pulsation, and magnetic field in a chemically peculiar environment makes the roAp stars important laboratories for testing stellar structure and evolution theory.  
 
The star \ac\   is visible to the naked eye ($V=3.2$\,mag). It   is a well-known and well-observed member of the roAp  group,and has a mean quadratic magnetic field of 4\,kG to 7\,kG \citep{mathys3},  a main pulsation period of 6.8\,min, and a rotation period of  about \prot\,=\,4.48\,d. \citet[][hereafter Paper\,I]{paper1} have already summarised past investigations of \ac , provided references, and discussed striking differences between the rotation light curves deduced from red and blue filter \bc\ observations\  \citep{brite1}. 

The  data needed for  clarification were obtained by three of the five \bc\ satellites in 2016. In addition we analysed \w\ data from 2006 and 2007 \citep{bruntt2}, and data from  \s\ obtained from 2003 to 2010 \citep{jack,tarr} which cover the two last \w\ seasons. Finally, we also address  \t\ data \citep{tess}  obtained in 2019. 

The large time base (almost 20 years of observing \ac\ with various space missions) allows us to discuss rotation and pulsation, with a discussion of the inherent problems of photometric surface imaging using Bayesian techniques, which has two main advantages: (i) parameter estimation with the Bayesian technique provides credibility regions from the data alone and (ii) it allows us to rank models according to their evidence.

Concerning pulsation properties, the dependence of amplitude and phase of the dominant frequency f$_1$ on the rotation phase of \ac \ (see Paper\,I) was another challenge for  the 2016 BRITE photometry. Merging the results with space photometry obtained by \w , \s , and \t\ helped to improve the accuracy significantly. With a much better signal-to-noise ratio (S/N) of the new red and blue 2016 BRITE data, we also address  other frequencies mentioned by \citet{bruntt2} and in Paper\,I.

\section{Photometric data and reduction}

As \ac\ was observed by different satellites at different epochs, the various data sets exhibit individual peculiarities. 
The effective wavelengths of the  passbands of the various satellites (Table\,\ref{t:pass}) were determined by using the  filter values and  a synthetic spectrum with \Teff\, =\,7\,500\,K, \lg\,=\,4.1\,\cms,\  and  a chemical composition that agree  with the atmospheric parameters determined by  \citet{koch}  for \ac .

\begin{table}[h]
\caption{Effective wavelengths and band widths of the space photometers \bab\ (BRITE-Austria), \blb\ (BRITE-Lem), \btr\ (BRITE-Toronto), SMEI, WIRE, and TESS. The combined \bab\ and \blb\ data are indicated in the text as Bb$^{\star}$.
        }              
\label{t:pass}       
\centering                           
\begin{tabular}{lrr}       
\hline\hline                  
            & $\lambda$ (nm) & band width (nm) \\ 
\hline                        
   \bab\  \&\   \blb\ (= Bb$^{\star}$)  &  425                  & 55 \\
    \btr                     &  605                  & 145 \\
   WIRE                 &  $\approx$ 600  &  $\approx$ 380\\  
   SMEI                  &  630                   & $\approx$ 600 \\
   TESS                 &  $\approx$ 800   & $\approx$ 400 \\
\hline                                   
\end{tabular}
\end{table}

\subsection{BRITE-Constellation}

\ac\ was observed during the commissioning phase of \bc\   \citep{brite1} from March 3, to August 8, 2014, for 146 days. The analysis of these data is presented in Paper\,I, which describes a first attempt at photometric surface imaging of \ac\  in blue and red colours, based on Bayesian techniques and using the rotation period of 4.4790\,days, determined by \citet{kurtz2}. 

The star was observed again in the 15-CruCar-I-2016 field from February 4, to July 22, 2016, 
by three of the five operational nanosatellites, which provided a  total of 163\,433 photometric measurements from \btr, \bab, and \blb\ (Table\,\ref{t:sum})  and which are the basis for the present investigation. 
Each satellite obtained 10 to 30 measurements per satellite orbit (about 101min) with a typical sampling of 20\,s and exposure time of 1\,s.

The pipeline outlined by \citet{po1} and \citet{po2} was used to process the raw images from the satellites. While the pipeline accounts for technical issues typical of the BRITE photometry, such as hot pixels \citep{pa}, the extracted photometry remains affected by systematic instrumental effects. These systematics, such as CCD temperature drifts and the position of the star's point spread function (PSF) in the raster, are identified and removed via decorrelation using a procedure similar to that outlined by \citet{pig} and \citet{pig2}. An additional decorrelation following  the procedure outlined by \citet{buy} mitigates the impact of the PSF modulation with temperature. 

\begin{table}[h]
\caption{Observations of \ac\ obtained in 2016 by \bab\ (BRITE-Austria), \blb\ (BRITE-Lem), and \btr\ (BRITE-Toronto).  
        }              
\label{t:sum}       
\centering                           
\begin{tabular}{l c c r r}       
\hline\hline                 
BRITE & filter & number  of  &\multicolumn{1}{c}{start} & \multicolumn{1}{c}{end}  \\    
            &        & data points    &\multicolumn{2}{c}{dd.mm.yyyy} \\
\hline                        
   \bab  & blue & 29395 &  04.02.2016 &  27.05.2016\\
   \blb   & blue & 48522 &  03.03.2016 &  15.07.2016\\
   \btr    &  red & 85516 &  11.02.2016 &  22.07.2016\\
\hline                                   
\end{tabular}
\end{table}

\subsection{WIRE}

The Wide Field Infrared Explorer (WIRE) was launched on March 4, 1999, but the hydrogen cryogen boiled off soon after launch due to a technical problem, which terminated the primary science mission. The onboard star tracker with 52\,mm aperture, however, remained functional and could be used for long-term visual precision photometry until communication with the satellite failed on October 23, 2006 \citep{buz,bruntt3,bruntt4}.
Our target, \ac, was observed in September 2000,  February 2005, and February and July 2006, for a total of 84 days \citep{bruntt2}, of which we are addressing the 2005 and 2006 data (February 16, 2005, to August 1, 2006).

\subsection{SMEI}

The Solar Mass Ejection Imager (SMEI) is an instrument on board the Coriolis satellite, which was launched on January 1, 2003. The primary science goal is to detect disturbances in the solar wind, but in doing this the three CCD cameras observed  the whole sky in successive passes. These data were used to detect, among other things, stellar pulsation \citep{jack,tarr,Hounsel}. Our target star \ac\ was observed by SMEI from February 3, 2003, to December 30, 2010. 

Raw SMEI data suffer from very strong instrumental effects. The final light curve was obtained by correcting first for a one-year periodicity, then detrending and sigma clipping, which was repeated up to 25 times. Finally, signals in the vicinity of 1, 2, .... 6 \cd\ were subtracted. The frequency spectrum of the final data has a sharp decrease at the lowest frequencies, which is the result of detrending. A  signal at the satellite orbital frequency, its multiples, and side lobes at typically $\pm$1 and $\pm 2\,{\rm d}^{-1}$ were also removed. 
Unfortunately, we do not know all the details of the SMEI photometry. A single SMEI data point comes from a series of 4 s exposures, but it is not always clear how many individual exposures are combined. This depends on the time a star passes through the camera field of view (3\degr$\,\times\,$60\degr ), and changes with the aspect angle. Thus, the total integration time is typically  below one minute, but can sometimes be slightly longer.

\subsection{TESS}

TESS was launched on April 18, 2018, and has four identical wide-field cameras that together monitor a $24\degr$ by $96\degr$ strip of the sky and with a red-optical passband. Each field is monitored for about 27 days.

The TESS data for \ac\ span from April 24, to June 18, 2019, in sectors 11 and 12 with a baseline of 55 days (four orbits of  13.7 days each) with a 2 min cadence. 
We downloaded the Simple Aperture Photometry from the MAST portal\footnote{\url{https://mast.stsci.edu/portal/Mashup/Clients/Mast/Portal.html}}, applied a 3$\sigma$ outlier clipping (relative to a 2-day moving average), and used the data without any further corrections. 

\begin{table*}[h]
\caption{Estimated parameter values for spot models based on BTr, \w , \s, and \t\ data.}
\label{t:BWST}
\centering
\renewcommand{\arraystretch}{1.5}
\begin{tabular}{l r r r r r r r r r}
\hline\hline
         & \multicolumn{2}{c}{BTr}  & \multicolumn{2}{c}{WIRE}  & \multicolumn{2}{c}{SMEI}  & \multicolumn{2}{c}{TESS} \\
\multicolumn{1}{r}{ }    & mean & mode  & mean & mode  & mean & mode  & mean &  mode &    \\
\hline
\ \ \ spot\,1: \\
longitude (\deg)       & 0$^{+ 4}_{- 4}$  & 9  &    0$^{+2}_{-2}$ &    0 &   0$^{+7}_{-7}$  & 358 &  0.0$^{+0.3}_{-0.3}$   &   0.0 \\%
latitude  (\deg)       & -27$^{+14}_{-17}$  & -63 & 37$^{+5}_{-5}$ &  43 & -19$^{+16}_{-28}$    & -10 & -2.7$^{+0.6}_{-0.5}$   &  -2.4 \\%
radius    (\deg)   &  21$^{+12}_{- 9}$  &  52 & 14$^{+3}_{-3}$ &  17 &  19$^{+20}_{-9}$     &  14 &  9.7$^{+0.1}_{-0.1}$   &   9.7 \\%
\hline
\ \ \ spot\,2: \\
longitude (\deg)        & 171$^{+ 4}_{- 5}$  & 183 &179$^{+3}_{-4}$ & 178 & 155 $^{+7}_{-7}$     & 155& 144.4$^{+0.5}_{-0.5}$  & 144.5 \\%
latitude  (\deg)       & -30$^{+12}_{-12}$ & -37  & 40$^{+6}_{-6}$  &  44  &  46  $^{+25}_{-23}$ &  67 &  20.7$^{+0.6}_{-0.6}$   &  20.9 \\%
radius    (\deg)   &  21$^{+ 9}_{- 8}$   &  24  & 13$^{+3}_{-3}$  &  17  &   8   $^{+2}_{-2}$     &  10 &    6.7$^{+0.1}_{-0.1}$   &    6.7 \\%
\hline
\ \ \ spot\,3: \\
longitude (\deg)       & 111$^{+20}_{-20}$  & 133 &120$^{+2}_{-2}$ & 120 & -- & -- & 171.9$^{+0.6}_{-0.6}$  & 171.7 \\%
latitude  (\deg)      &  34$^{+29}_{-22}$   &  -4   &-10$^{+2}_{-2}$  & -11 & -- & -- & -36.1$^{+5.3}_{-5.1}$   & -36.8 \\%
radius    (\deg)  &    6$^{+ 1}_{- 1}$     &   9   &  9$^{+1}_{-1}$   &   9  & -- & -- &  16.9$^{+4.2}_{-4.3}$    &  17.5 \\%
\hline
period  (days) \vspace{-2mm}  & 4.4779           & 4.4781 & 4.47925            & 4.47926 & 4.47912            & 4.47912 & 4.4803           & 4.4803 \\
                                                 &$\pm 0.0012$ &             & $\pm 0.00009$ &               & $\pm 0.00018$ &               & $\pm 0.0004$   \\
residuals (mmag)  &$\pm 1.386$   & & $\pm 0.318$ &           & $\pm 6.232$      &               & $\pm 0.141$  \\
\hline
\end{tabular}
\tablefoot{Longitudes are with respect to HJD 2457510.5105 (BRITE-Toronto), HJD 2453680.6598 (WIRE), HJD 2454115.2887 (SMEI), and HJD 2458625.8020 (TESS). Spots are ordered according to their impact on the light curve. Modal values are provided together with mean values and 90\% credibility limits.
The set of mean parameter values may differ markedly from the set of modal values due to the skewness of the posterior. 
The residuals in mmag of the photometric data to the model light curve are r.m.s. values.
}
\end{table*}

\section{Bayesian photometric imaging}                                   \label{s:bayes}

Our Bayesian photometric imaging technique is described in detail by \citet{lueft} and in our Paper\,I. The number of free parameters $N$ depends on the complexity of the stellar surface model considered. The following spot parameters are estimated: longitude, latitude, and radius for each spot, which is assumed to be circular. Therefore, a three-spot model involves at least ten free parameters, including the rotation period. 

Published values were used for the following parameters: \\
$\bullet$\ \ Inclination ($i\,=\,36^\circ$). This value results from ${\rm v\,sin} i$ and the stellar radius given by \citet{bruntt1}. \\
$\bullet$\ \ Quadratic limb darkening ($U_{\rm a}=0.278, U_{\rm b}=0.382$). Limb darkening depends on wavelength, and also on  rotation phase because  \ac\ is a mCP star. We  chose representative values based on a model atmosphere with a \Teff\,=\,7\,500\,K and \lg\,=\,4.1\cms\ \citep{koch}. We note that the exact choice of the limb darkening coefficient is not critical here as it is degenerate with the spot size resulting from the light curve modelling. \\
$\bullet$\ \ Contrast between spot and  undisturbed photosphere. This has to be fixed because spot area and brightness contrast are highly anti-correlated. The spots are almost certainly bright in the optical wavelengths as the flux absorbed in the UV is redistributed in the red part of the spectrum for \Teff\ values typical for mCP stars (see Section 4.2. of Paper\,I). To start we assume a contrast of $\kappa\,=\,1.25$.

For our Bayesian surface imaging we use the average magnitude during each orbit of the respective BRITE satellite after carefully removing instrumental effects. This  binning ensures that the brightness variation due to stellar oscillations (less than 2\,mmag in the blue and even less in the red filter, and of the order of few minutes) cancels out. The much higher noise level of the BRITE-blue data\footnote{In the following we use  Bb$^{\star}$ for the combined  BRITE-blue (\bab\ and \blb\ ) data.}  (\bab\ and \blb ) compared to the red data (\btr ) is evident. Fortunately, the amplitudes are larger in the blue.

Similarly, the 36\,124 TESS data points were binned for our surface imaging into 4019 time bins, with a maximum bin size of 16.1 minutes. This results in up to nine original data points per bin with a typical scatter of $\pm$\,0.37\,mmag per data point. The accuracy of a bin-mean is up to three times better.

\subsection{Bayesian concept}   \label{s:concept}

In a nutshell, a Bayesian approach consists of the following: An uninformative {prior} probability distribution in the N-dimensional parameter space is converted by the likelihood of the data, given a set of parameter values, into a {posterior} probability distribution. This posterior contains what  can be learned from the data in terms of a given model. It is common practice to extract N marginal distributions, one for each parameter, from the posterior. Each marginal distribution can be summarised by a mean value and  a  90\% credibility interval. In a few cases we use a 68\% interval.

In a second step,    a model's {evidence} can be determined by integrating the posterior probability distribution over the parameter space and dividing by the volume of the latter because it is the mean probability that is important. This is computationally demanding, but allows one to  {quantitatively} compare different models (e.g.  a two-spot model versus a three-spot model).

The centre of gravity (the barycentre of the N-dimensional posterior \lq probability mountain\rq)  is defined by the N parameter mean (expectation) values, computed  as described above;  it minimises the mean quadratic dispersion.
It is this barycentre that makes best use of all the available information as less probable sets of 
parameter values also contribute to a model's evidence. This mean is more relevant than the locus of maximum  probability, the mode (best fit). Only in the rare case of a symmetric probability mountain do mean and mode coincide.

A corresponding picture could be a  comparison of  the height of the Matterhorn in Switzerland with that of  Table Mountain in Cape Town, South Africa, and quote their positions. The case of the  Matterhorn is trivial. It has a  nearly  symmetric structure and an obvious peak above the timberline, but Table Mountain raises a problem. 
Is the position of the mini-peak at the border of the Table (location of the \lq mode\rq ) to be listed, or is it more appropriate to quote the geographical centre (centroid, \lq mean\rq ) of the plateau?

Probabilistic methods do not primarily aim for a single best-fit solution of a given problem, but for the posterior distribution of the various model parameters and how well a model reproduces the observations {compared to other models}. However, for the sake of completeness we also provide modal values in Table \ref{t:BWST}, which summarises the model parameters derived from BTr, \w , \s , and \t\ observations.

     \begin{figure}[h]
      \resizebox{\hsize}{!}{\includegraphics{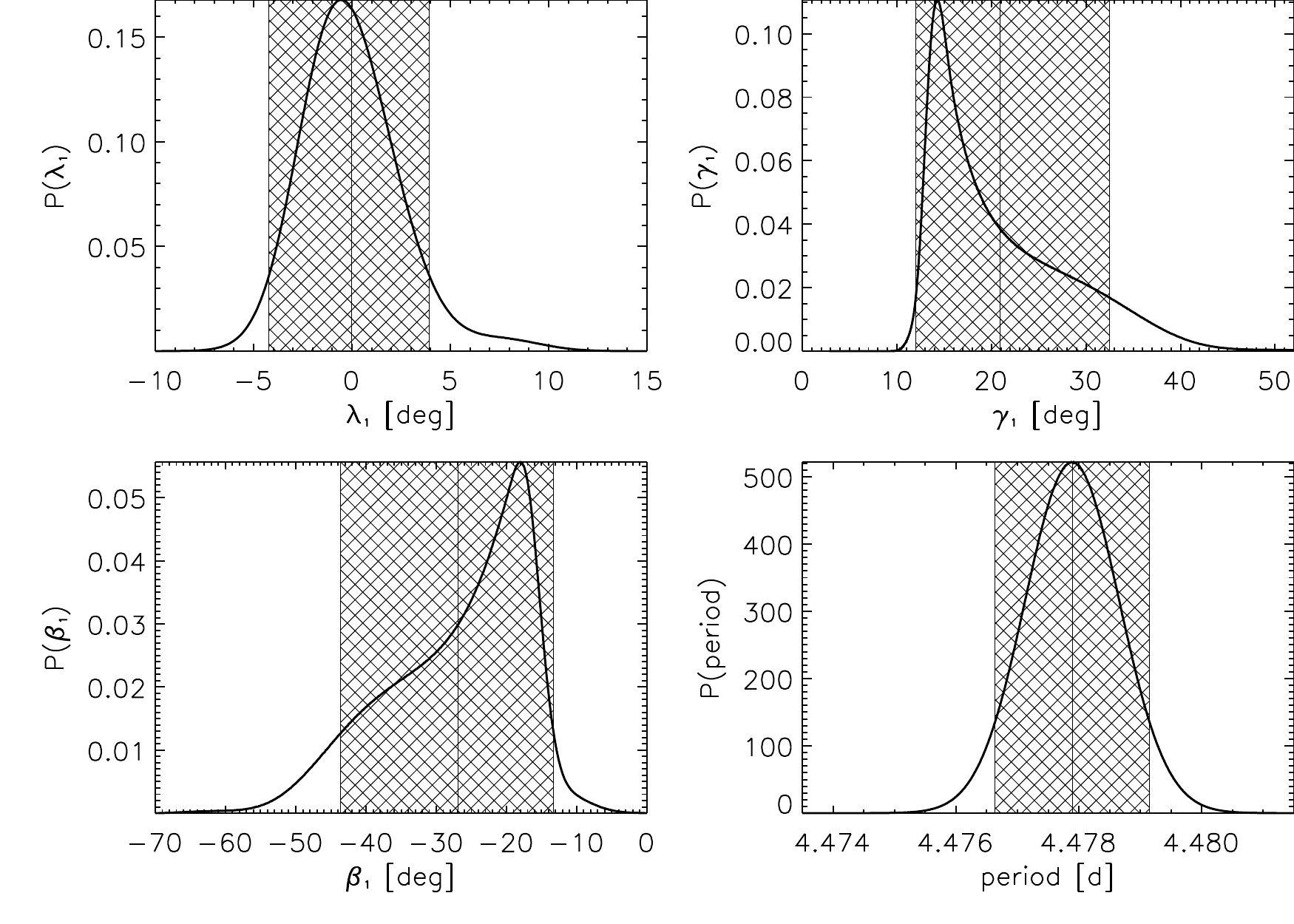}}
      \caption{Left: Spot\,1 
      marginal distributions of longitude ($\lambda$) and latitude ($\beta$) in degrees. Right: Spot radius ($\gamma $) and period in days. All for BTr data. The vertical lines indicate the mean and the 90\% credibility limits (shaded area). Because of the skewness of the distribution (curved line) the locus of maximum probability does not necessarily coincide with a parameter's mean value and deviates from the global mode given in Table\,\ref{t:BWST} for BTr.
                }  
         \label{f:mcmc}
   \end{figure}

Figure\,\ref{f:mcmc}, deduced from the BTr data, shows the marginal distributions for longitude, latitude, and radius of  spot 1,  which has the largest impact on the light curve, and the marginal distribution for the rotational period. 
Each marginal distribution results from integrating the $N$-dimensional posterior over the corresponding $(N-1)$-dimensional subspace. The boundaries of the 90\% credibility interval as well as the parameter mean are indicated by vertical lines. The widths and asymmetries of the marginal distribution for latitude and radius indicate that these two model parameters are poorly defined, whereas spot longitude and rotation period are much better constrained by the data. Because of the skewness of the distribution, the locus of maximum probability does not necessarily coincide with a parameter's mean value and may be very different from the global mode presented in Table\,\ref{t:BWST}.

It should be noted that finding the posterior's mode is computationally fast in comparison to the Markov chain 
Monte Carlo (MCMC) integrations, which are necessary to provide mean values and credibility intervals.

\subsection{Rotation}  \label{s:rot}        

An early estimate for a rotation period, based on magnetic field measurements  \citep{wood} and $H_{\alpha}$ polarimetry \citep{land} is given by \citet{mathys2}, who speculated that the star has a magnetic field that varies with a period of more than two weeks. Various photometric, spectroscopic, and magnetic field investigations of \ac\ were published in the following years claiming a rotation period between 4.46\,d and 4.48\,d (\citet{kurtz2}, \citet{balona}, \citet{bych}, \citet{bruntt2}, \cite{hubrig}).

   \begin{figure}[h]
      \includegraphics[width=\hsize]{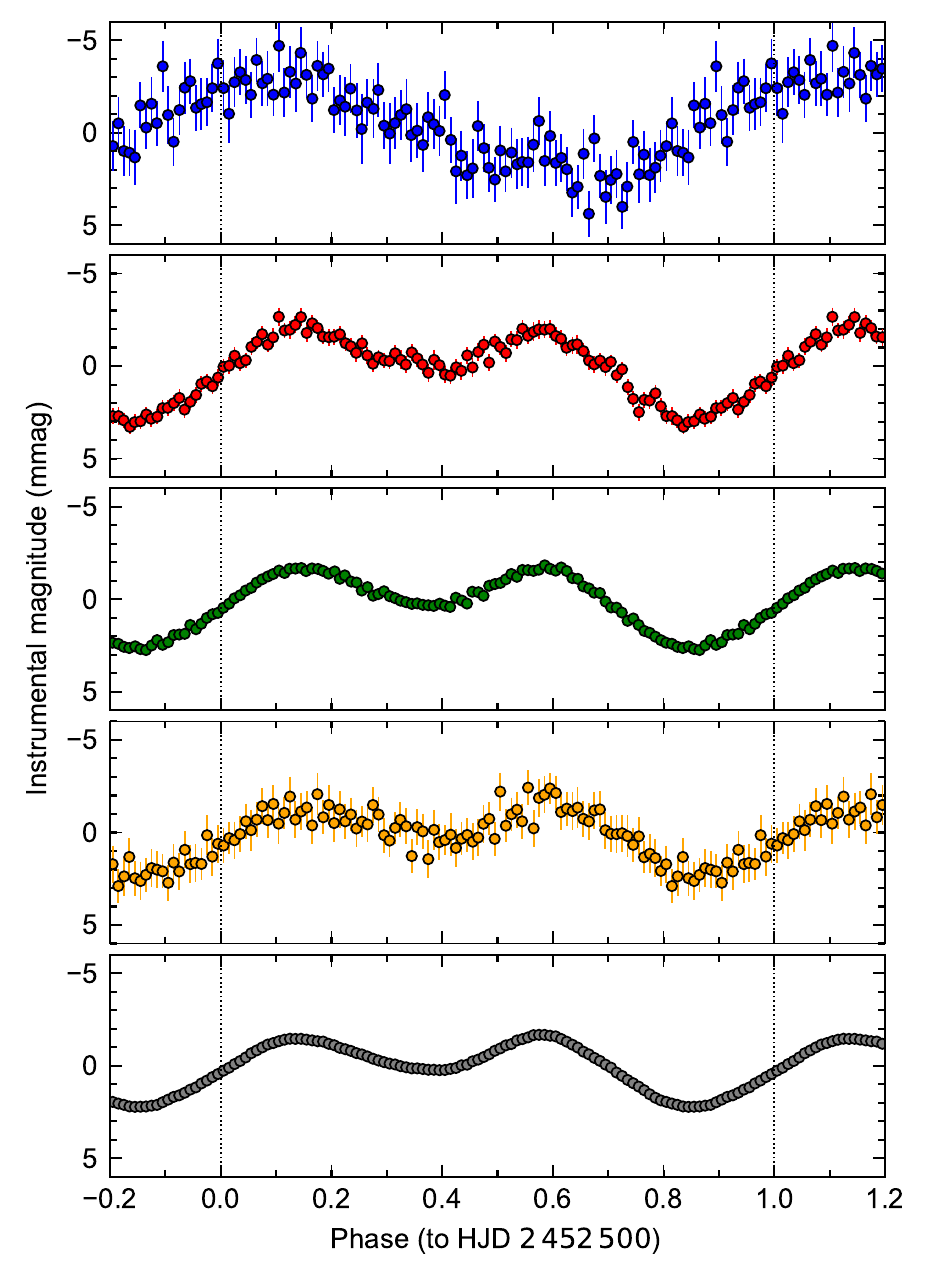}
      \caption{Light curves of \ac\ folded with the rotation period \makebox{$P_{\rm rot}\,=\,4.47930$\,d} and for the epoch of HJD\,=\,2\,452\,500. The panel sequence follows the central wavelength of the used filters (Table\,\ref{t:pass}), and  ranges from 425\,nm (Bb$^*$), top panel, and about 600\,nm (BTr, WIRE, SMEI), to 800\,nm (TESS), bottom panel.
      }
      \label{f:rot}
  \end{figure}

Figure\,\ref{f:rot} shows the rotational light curves of \ac\ observed by six satellites. The differences between the shapes of the blue (Bb$^*$) and  red light curves indicate a complex balance between line or continuum opacities, which removes flux from the blue band, 
but increases flux at other wavelength regions due to back-warming, as is discussed by   \citet{dleck} and \citet{denis}, among others. In particular, surface spots with different chemical compositions, typical of CP stars, can cause complex effects on light curves obtained with different filters. Extended spectroscopic investigations are needed to explore such a scenario.  

The stellar rotation period is observationally the most obvious parameter and can be estimated, in principle, without a model. However, here we consider the rotation period as an unknown parameter of a three-spot model (Table\,\ref{t:BWST}).
A re-analysis of the  BRITE-red 2014 data from Paper\,I, with the rotation period being a further free parameter in the Bayesian analysis, indicates a \prot\,=\,4.4846\,$\pm 0.0017$\,d (68\% interval), which exceeds the standard value of 4.4790\,d by 3.3\,$\sigma$.  

\begin{figure*}[h]
     \center\includegraphics[width=0.9\hsize]{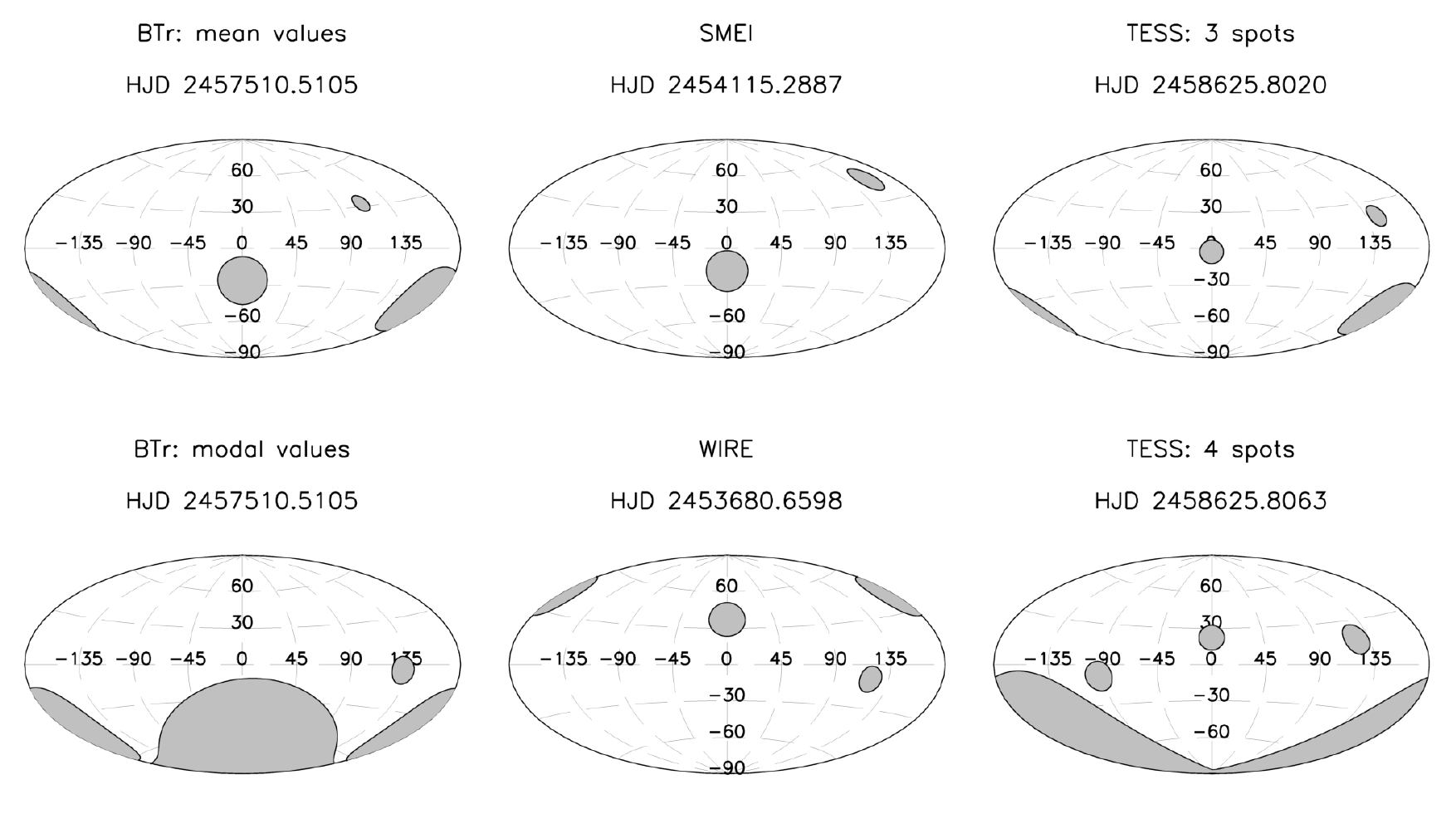}
      \caption{HJDs are transit times of the spot that contributes most to the light curve (spot\,1). Left: BTr-based maps (top: mean values; bottom: modal values). Middle: SMEI mean map, which allows  for only two spots (top) and WIRE (bottom), which indicates a three-spot map. Right: Comparison of TESS mean maps allowing for three spots (top) and   four spots (bottom).  
                                 }
     \label{f:allmaps}
   \end{figure*}

A comparison of spot\,1\ transit times from WIRE and TESS (three-spot model) resulted formally in \prot\,=\,4.47930\,$\pm 0.00002$\,d (68\% interval), assuming that there are 1104 stellar revolutions between the corresponding maps (Fig.\,\ref{f:allmaps}).

\subsection{Spots}      \label{s:spot}          

A major challenge is estimating the number of spots necessary to get a good fit without overfitting. In Paper\,I we were able to identify one bright spot in the \b -blue data, two spots in \b -red and WIRE data, but there were already hints of a more complex surface map, at least in the red case. Not surprisingly, the photometric quality of the data matters. In contrast to the 2014 data, the 2016 BTr data convincingly indicate a three-spot model. The reduction in credibility, which is a formal consequence of an increased number of free parameters from seven (two spots) to ten (three spots) is more than compensated for by an improved goodness-of-fit value.

With the MCMC technique and the availability of PC clusters, an integration over a high-dimensional configuration space is meanwhile computationally feasible. With computed evidence it is possible  to  quantitatively compare the performance of models differing in the number of spots by considering the evidence ratio (i.e. the Bayes factor). To our knowledge, this is the first time that the evidence of models differing in the number of spots has been probabilistically determined. 
The poor quality of the  \b-blue\ photometry prevented an independent estimate of \prot\  as a Bayesian spot modelling parameter.

\subsubsection{Red maps}

For the red data  obtained with \btr , the evidence gain from a two-spot model to a three-spot model is approximately $1.2\times 10^{6}$. This large Bayes factor is due to the  number of data points (1678) involved.  The residuals to the photometry decreased slightly from 1.389 to 1.386\,mmag and the mean gain of evidence {per data point} is 0.8\%\ ($(1.2\times 10^{6})^{1/1678}=1.008$) in favour of the three-spot model. 
Table\,\ref{t:BWST} summarises our three-spot model MCMC calculations based on BTr, \w , \s, and \t\ photometry. The resulting maps are illustrated in Fig.\,\ref{f:allmaps}. 
The tenth parameter listed in  Table\,\ref{t:BWST} is the rotation period with 90\%\ credibility ranges. We note  that the usual 68\% interval for \prot\ is (for a Gaussian distribution) smaller by a factor of 1.645 (e.g. for BTr: $\pm\,0.0008$\,d). The BTr maps are based on 62735 \btr\ input data points obtained in 2016, and binned in 1678 time bins according to individual BRITE orbits, and are shown in the left column of Fig.\,\ref{f:allmaps}. 

The rather poor photometric quality of the \s\ data does not allow us to obtain a three-spot model because the MCMC algorithm simply does not converge; only two spots can be identified (Fig.\,\ref{f:allmaps}, top in middle column).
The surface map based on nearly 55 days of TESS (2019) photometry (Fig.\,\ref{f:allmaps}, right column), confirms   the map derived from BTr (2016) data. Overall, the TESS data quality is impressive and encourages   looking for a fourth spot (cf. Table \ref{t:TESS_4spots:mcmc}, and Fig.\,\ref{f:allmaps}). The addition of another spot is rewarded by a substantial gain in goodness of fit. The residuals drop from \mbox{$\pm$0.141\,mmag} (3-spot model) to $\pm$0.125\,mmag. 

\begin{figure}[h]
     \center\includegraphics[width=0.9\hsize]{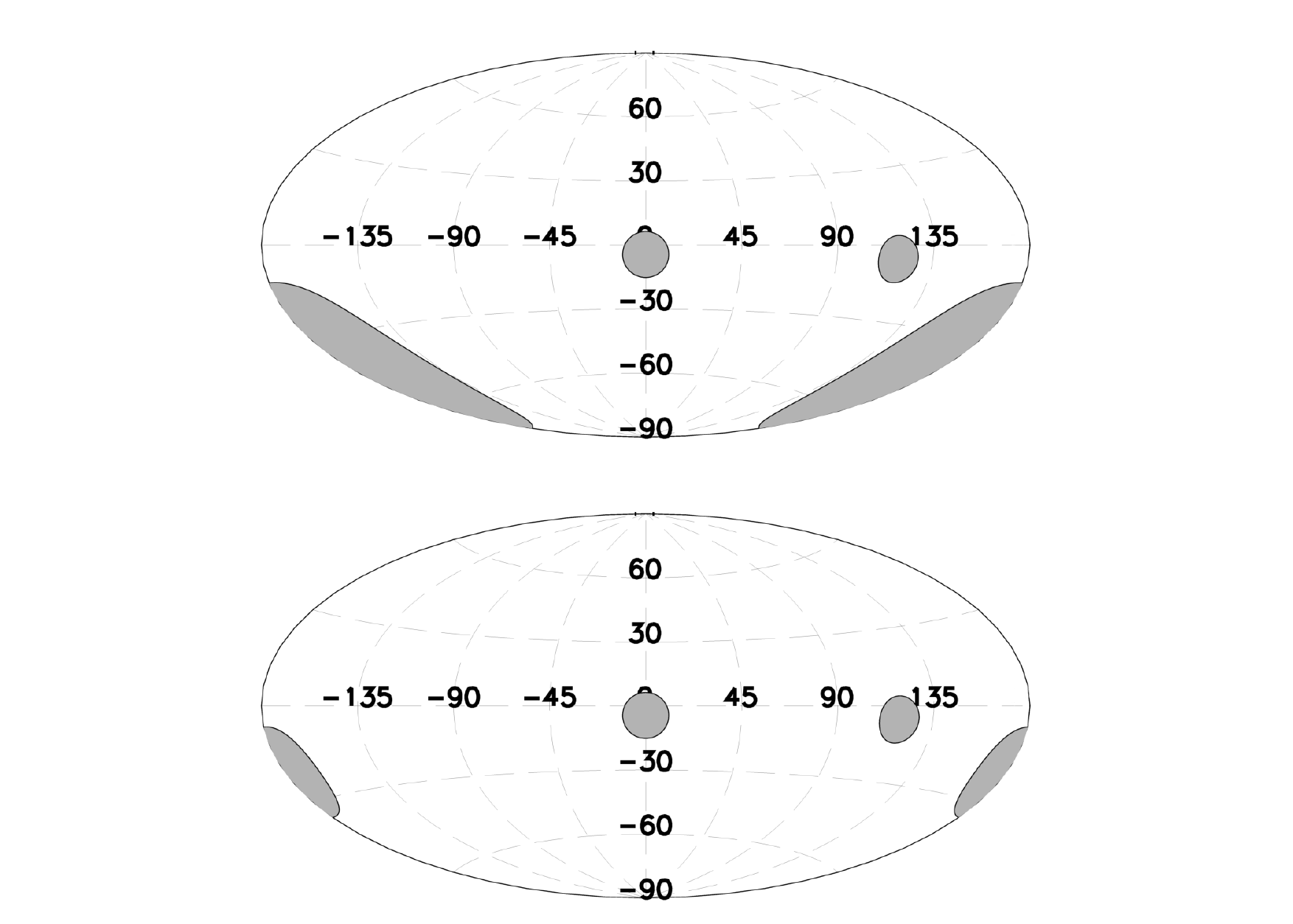}
     \caption{Two maps derived from the same WIRE data. Both maps fit the light curve with
an r.m.s. error only 0.002\,mmag larger than the WIRE map presented in Fig.\,\ref{f:allmaps}.
     }
     \label{f:Wvar}
\end{figure}

A comparison of  the maps in Fig.\,\ref{f:allmaps}, middle column, on top (\s ) and at the bottom (\w ), seem to imply that the orientation of one map is flipped relative to the equator. This illustrates the difficulties one encounters when trying to convert a one-dimensional light curve into a two-dimensional surface map. While  period and spot longitudes are comparatively well defined, estimating spot latitudes notoriously proves to be ill-posed unless the S/N is very high. Figure\,\ref{f:Wvar} illustrates the problem, showing two alternative \w\ maps that are only marginally inferior to the solution shown in Fig.\,\ref{f:allmaps}, but with spot latitudes comparable to maps from BRITE and TESS data. 

\begin{table}[h]
\caption{Parameter values for our four-spot model based on binned TESS data (4019 points) obtained in 2019 .}
\label{t:TESS_4spots:mcmc}
\centering
\renewcommand{\arraystretch}{1.5}
\begin{tabular}{l r r}
\hline\hline
           &\multicolumn{2}{c}{\t } \\
           & mean  &  mode   \\
\hline
spot\ 1: longitude      (\deg) &  0.0$^{+1.0}_{-1.0}$   &  -0.2 \\%
\hspace{10mm} latitude  (\deg) & 22.1$^{+1.1}_{-1.1}$   &  22.2 \\%
\hspace{10mm} radius    (\deg) & 10.3$^{+0.8}_{-0.9}$   &  10.5 \\%
\hline
spot\ 2: longitude      (\deg) & 124.0$^{+1.7}_{-1.6}$  & 123.8 \\%
\hspace{10mm} latitude  (\deg) &  17.2$^{+3.4}_{-3.5}$  &  18.2 \\%
\hspace{10mm} radius    (\deg) &  10.1$^{+0.5}_{-0.5}$  &  10.2 \\%
\hline
spot\ 3: longitude      (\deg) & 198.8$^{+1.4}_{-1.5}$  & 199.3 \\%
\hspace{10mm} latitude  (\deg) & -48.8$^{+3.3}_{-3.4}$  & -48.0 \\%
\hspace{10mm} radius    (\deg) &  45.0$^{+1.7}_{-1.7}$  &  44.6 \\%
\hline
spot\ 4: longitude      (\deg) & 265.9$^{+0.9}_{-0.9}$  & 266.0 \\%
\hspace{10mm} latitude  (\deg) &  -8.6$^{+0.8}_{-0.8}$  &  -8.4 \\%
\hspace{10mm} radius    (\deg) &  11.0$^{+1.7}_{-1.8}$  &  11.5 \\%
\hline
period (days)            & 4.4799$^{+0.0003}_{-0.0003}$ & 4.4799\\%
residuals (r.m.s. in mmag)  & $\pm 0.125$  \\ 
\hline
\end{tabular}
\tablefoot{Spots are ordered according to their impact on the light curve. 90\% credibility limits are given to the values in the column \lq mean\rq . Spot longitudes are with respect to HJD 2458625.8063 (transit time of spot 1, see also Fig.\,\ref{f:allmaps}, bottom of right column).}
\end{table}

\subsubsection{Blue maps}        \label{s:bmap}

In the next step we reanalysed the much noisier blue data obtained with Bb$^{\star}$. 
A reduced data quality affects  the sophistication of models that can realistically   be tested, and   is  expressed in the number of free parameters.  For the noisy Bb$^{\star}$ data we limited this number as much as possible by fixing all but one parameter, the brightness contrast (higher--lower\,: $\kappa$ is larger--smaller than 1), and being the same for all three spots.  
We found $\kappa = 1.05\pm 0.04$, albeit with residuals as large as $\pm 5.9$\,mmag (r.m.s.), when applying our red model ($\kappa = 1.25$) to the blue data. We recall that the blue photometry is about a factor of four noisier than the red photometry (Fig.\,\ref{f:rot} and Paper\,I). 

When we increase  the parameter space for a test  by assigning each spot an individual $\kappa$,  the blue spot 1 completely disappears, and hence the solution converges to a two-spot model. Instead,  when   we  formally allow   a time lag  $\Delta t$ in the three-spot model, the fit  improves: $\kappa = 1.23\pm\,0.035$, and $\Delta t = 0.69\pm\,0.08$\,d, with residuals of $\pm\,5.1$\,mmag (r.m.s.). The time lag of $\approx$\,17 hours is equivalent to a phase shift of $56^\circ\pm\,7^\circ$. 

Both of these models have  comparable Bayesian evidence, but whereas spots with different chemical compositions, and hence different $\kappa$ values, are known for CP stars, a colour-driven phase shift is not. In conclusion, the currently available blue photometry is insufficient for a reliable and convincing photometric surface mapping.

\begin{figure*}[h]
     \center\includegraphics[width=\hsize]{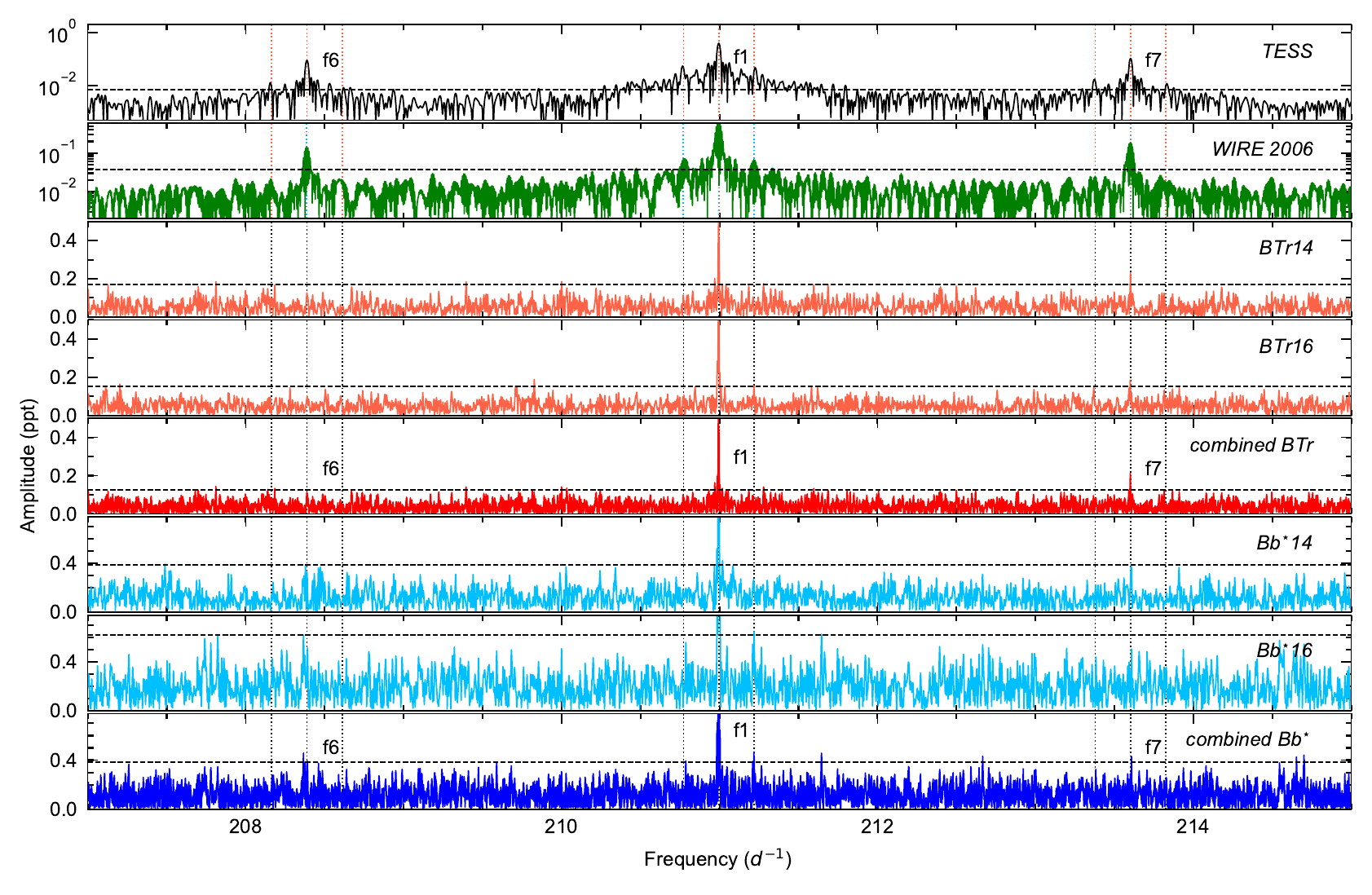}
     \caption{Fourier amplitude spectra from different sources around the primary pulsation period of \ac . Bb$^{\star}$ represents the combined blue data from BAb and BLb.  BTr14, BTr16, Bb$^{\star}$14, and Bb$^{\star}$16  are the red and blue BRITE data, obtained during the years 2014 and 2016, respectively. Vertical dotted lines indicate the oscillation modes and the rotational  side lobes identified in the TESS data. Horizontal dashed lines mark a S/N\,=\,3. We note that  the two upper  panels have a logarithmic scaling and that many peaks formally exceed the S/N=3 limit in the TESS data,  which are caused by the spectral window, as is typical for high S/N data. The y-scale of the other panels, however, is chosen to distinguish the noise, $f_6$ and $f_7$.}  
     \label{f:ap1}
\end{figure*}

\begin{figure*}[h]
     \includegraphics[width=\hsize]{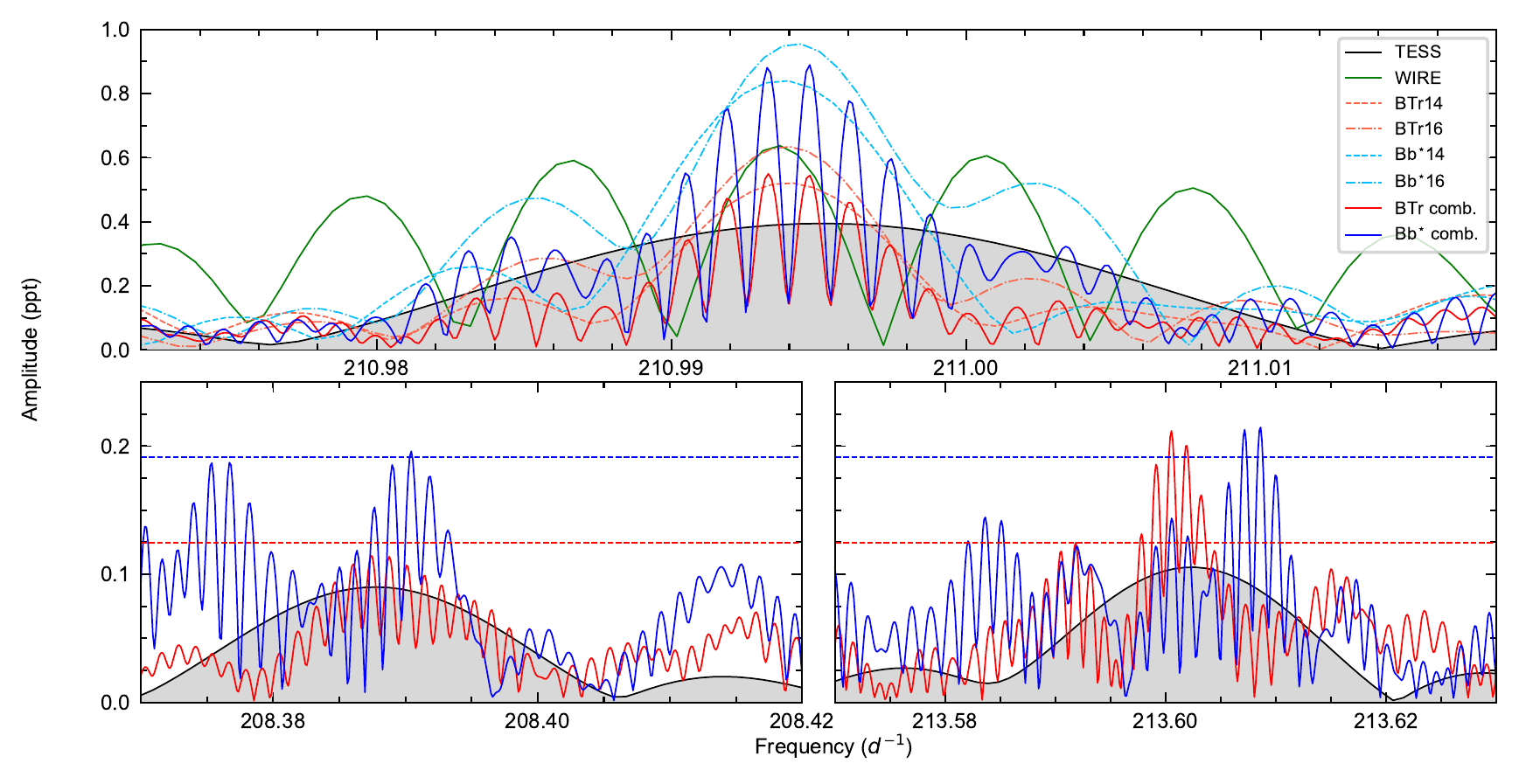}
     \caption{Fourier amplitude spectra of the TESS and BRITE photometry centred on $f_1$ (top), $f_6$ (bottom left), and $f_7$ (bottom right).  Bb$^{\star}$ represents the combined blue data obtained with BAb and BLb.  BTr14, BTr16, Bb$^{\star}$14, and Bb$^{\star}$16  are the red and blue BRITE data, obtained during the years 2014 and 2016, respectively. The  BRITE-blue spectra amplitudes are divided by two for better comparison with the other data. Dashed lines indicate a S/N of three. }  
     \label{f:ap2}
\end{figure*}

\section{Pulsation of \ac}         \label{s:puls}    

In the following we use the same numbering of frequencies as was used by \citet{bruntt2} and  in  Paper\,I.

\begin{table}[h]
\caption{Detected frequencies in the various time series of \ac .}
\label{t:freq}
\centering
\renewcommand{\arraystretch}{1.5}
\begin{tabular}{r l r r r}
\hline\hline
 & Frequency  & Amplitude & Phase   & $p$ \\
& [d$^{-1}$]     & [parts per & [degrees] &    \vspace{-2mm}      \\
&                      & thousand] & \\
\hline
\multicolumn{5}{c}{$f_1$} \\
TESS & 210.99512(15)& 0.394(8)& - & 1.0\\
WIRE & 210.99361(5) & 0.65(1) & - & 1.0\\
BTr & 210.99329(2) & 0.577(28) & 95(1) & 1.0\\ 
Bb$^{\star}$ & 210.99327(2) & 1.716(81) & 84(1)& 1.0\\  
\hline
\multicolumn{5}{c}{$f_{1-} = f_1 - f_\mathrm{rot}$} \\
TESS & 210.7716(6) & 0.049(3) & - & 1.0 \\
WIRE & 210.7702(1) & 0.083(4) & - & 1.0\\
Bb$^{\star}$  & 210.7866(2) & 0.263(86) & - & 0.31\\
\hline
\multicolumn{5}{c}{$f_{1+} = f_1 + f_\mathrm{rot}$} \\
TESS & 211.2188(7) & 0.039(3) & - & 1.0\\
WIRE & 211.2171(1) & 0.074(4) & - & 1.0\\
BTb & 211.2167(1) & 0.334(86) & - & 0.82\\
\hline
\multicolumn{5}{c}{$f_6$} \\
TESS & 208.3879(3) & 0.089(3) & - & 1.0\\
WIRE & 208.3866(1) & 0.145(4) & - & 1.0\\
BTr & 208.3888(1) & 0.098(30) & 23(7) & 0.77\\  
Bb$^{\star}$  & 208.3890(1) & 0.318(82) & 11(5) & 0.91\\        
\hline
\multicolumn{5}{c}{$f_7$} \\
TESS & 213.6055(3) & 0.105(3) & - & 1.0\\
WIRE & 213.6004(1) & 0.186(4) & - & 1.0\\
BTr & 213.6005(1) & 0.195(29) & 92(4) & 0.99\\  
Bb$^{\star}$  & 213.6011(2) & 0.375(83) & 76(7) & 0.99\\        
\hline
\end{tabular}
\tablefoot{The Bb$^{\star}$  frequencies are derived from the combined  BRITE-blue observations (BAb and BLb), obtained in  2014 and 2016. Phases are defined for HJD = 2457200 and are given in degrees. The parameter $p$ gives the probability that a frequency is statistically significant compared to no signal (i.e. due to noise). }
\end{table}

No pulsation is detected in the SMEI data with an amplitude exceeding  0.33\,mmag, which corresponds to a S/N of 4.5. This is somewhat surprising given that the amplitude of the dominant  pulsation frequency $f_1$ in the combined 2014 and 2016 BRITE-red data, with a similar effective wavelength, is equal to 0.57\,mmag (Fig.\,\ref{f:ap1}).  The amplitude in the WIRE data is equal to 0.65\,mmag. However, the SMEI passband is very wide and corresponds to the sensitivity curve of a front-illuminated CCD, which ranges between 400 and 1000\,nm (Table\,\ref{t:pass}). Given different shapes, amplitudes, and phases of the rotational modulation of \ac\ at different wavelengths, it might well be that the pulsation signal has too small an amplitude in white-light SMEI data.  

After subtracting the rotational modulation from the (unbinned) original \t\ time series, the Fourier amplitude spectrum shows a rich pattern of pulsation modes in the vicinity of $f_1$. However, a detailed analysis of this is beyond the scope of the present paper and we instead refer to a follow-up study that will present a detailed asteroseismic analysis (Kallinger et al., in preparation). Here, we use the oscillations in the TESS data primarily to guide and verify the analysis of the BRITE observations. For comparison we also show in Figs.\,\ref{f:ap1} and \ref{f:ap2} the Fourier spectrum of the combined 2006 WIRE data ($\sim$170\,d with a 114\,d gap in between).

\subsection{Primary pulsation frequency $f_1$ }     \label{ss.f1}

After subtracting the rotational modulation via spline fits in the rotation-phase domain, the primary pulsation frequency ($f_1$) at about 210.99\,\cd\ is easily detected in all four BRITE data sets (Fig.\,\ref{f:ap1}), which we  abbreviate as Bb$^{\star}14$ and  Bb$^{\star}16$ (all blue filter data from  2014 and 2016) and BTr14  and BTr16 (all red filter data from  2014 and 2016). Combining the 2014 and 2016 data results in a total time base of  $\sim$840\,d with a coverage of about 35\%. Merging the data improves the frequency resolution as well as the S/N, but as can be seen in Fig.\,\ref{f:ap2} the $\sim$540-day  gap between the two data sets causes strong aliasing with about $\pm0.0019$\cd\ and multiples. 

To extract the oscillation parameters from the various light curves, we use a probabilistic approach presented by \cite{kal2017a}. The automated Bayesian algorithm was originally developed to deal with multiple frequencies within the formal frequency resolution \citep{kal2017} but works with a mono-periodic signal  (within one formal frequency resolution bin) as well. The method uses the nested sampling algorithm {\sc MultiNest} \citep{feroz2009} to search for periodic signals in time series data and tests their statistical significance (i.e. not  due to noise) by comparison with a constant signal. A solution is considered real\footnote{According to the convention established by \citet{jeff1961}, the evidence for or against one of two hypotheses is considered `substantial' for $p\gtrsim0.75$, `strong' for $p\gtrsim0.91$, and `very strong' for $p\gtrsim0.97$.} if its probability $p = z_\mathrm{signal} / (z_\mathrm{signal} + z_\mathrm{noise}) > 0.9$, where $z$ is the global evidence\footnote{The global evidence is a normalised logarithmic probability describing how well the model fits the data with respect to the uncertainties, parameter ranges, and the complexity of the fitted model.} delivered by {\sc MultiNest}. 

Figure\,\ref{f:ap2} shows highly significant peaks at about 210.9933\cd\ in the combined BTr and Bb$^{\star}$  data, which  compare  well visually to those found in the WIRE and TESS data. We note that the  BRITE-blue  spectra amplitudes are divided by two in this figure for better comparison with the other data. The best-fit solutions are listed in Table\,\ref{t:freq}. The uncertainties might appear unrealistically small, especially for the TESS data with a time base of less than 1/15 of the combined BRITE data. However, the frequency uncertainty is also influenced by the S/N of a frequency \citep[e.g.][]{kal2008}, which is about 18 times better in the TESS data than in the BRITE data. 

While the frequencies extracted from the red and blue BRITE data agree exceptionally well with each other (within 0.00002\,\cd\ or 30\,$\mu$s, given a period of 409.4917\,s), there are small but significant differences to the frequencies found in the WIRE (0.0003\,\cd ) and TESS (0.0018\,\cd ) data. Such offsets might indicate a variable $f_1$, for example due to a companion or evolutionary effects, but  aliasing in the BRITE data could also  explain this (at least partly). Even though the strongest peaks in the BTr and Bb$^{\star}$  data are nearly identical (Fig.\,\ref{f:ap2}), it might be that the physically relevant peak is one of the neighbouring alias peaks. If we consider the alias at $+0.0019$\cd\ as the true frequency then it would almost perfectly resemble the TESS result. However, forcing our frequency analysis algorithm to fit this peak and comparing the resulting global evidence to the original one gives a probability of 0.97 {against} this scenario. 

In Table\,\ref{t:freq} we also provide the phase of $f_1$ in the blue and red filter, and determine a phase difference of 
$11\pm2\degr$, which is consistent with the phase lags between Johnson B and V of \makebox{$7.4\pm5.1\degr$} derived by \citet{weiss}, \citet{kurtz1}, and the value given in Paper\,I ($10.6\pm5.9\degr$). Unfortunately, we cannot compare this result to the phases of the WIRE and TESS data. The slightly different frequencies we find in the data cause cumulative phase shifts (relative to the time of the BRITE observations) of more than 2.6 and 1.1 oscillation cycles for TESS and WIRE, respectively, so that comparing any phases becomes meaningless.

In conclusion, the average value for $f_1$, based on {all} BRITE data (years and colours) is 210.99328(2)\,\cd . In addition, we find evidence that this frequency changes over time, but we leave a more detailed discussion to a follow-up study. 

\begin{figure}[h]
     \center\includegraphics[width=0.9\hsize]{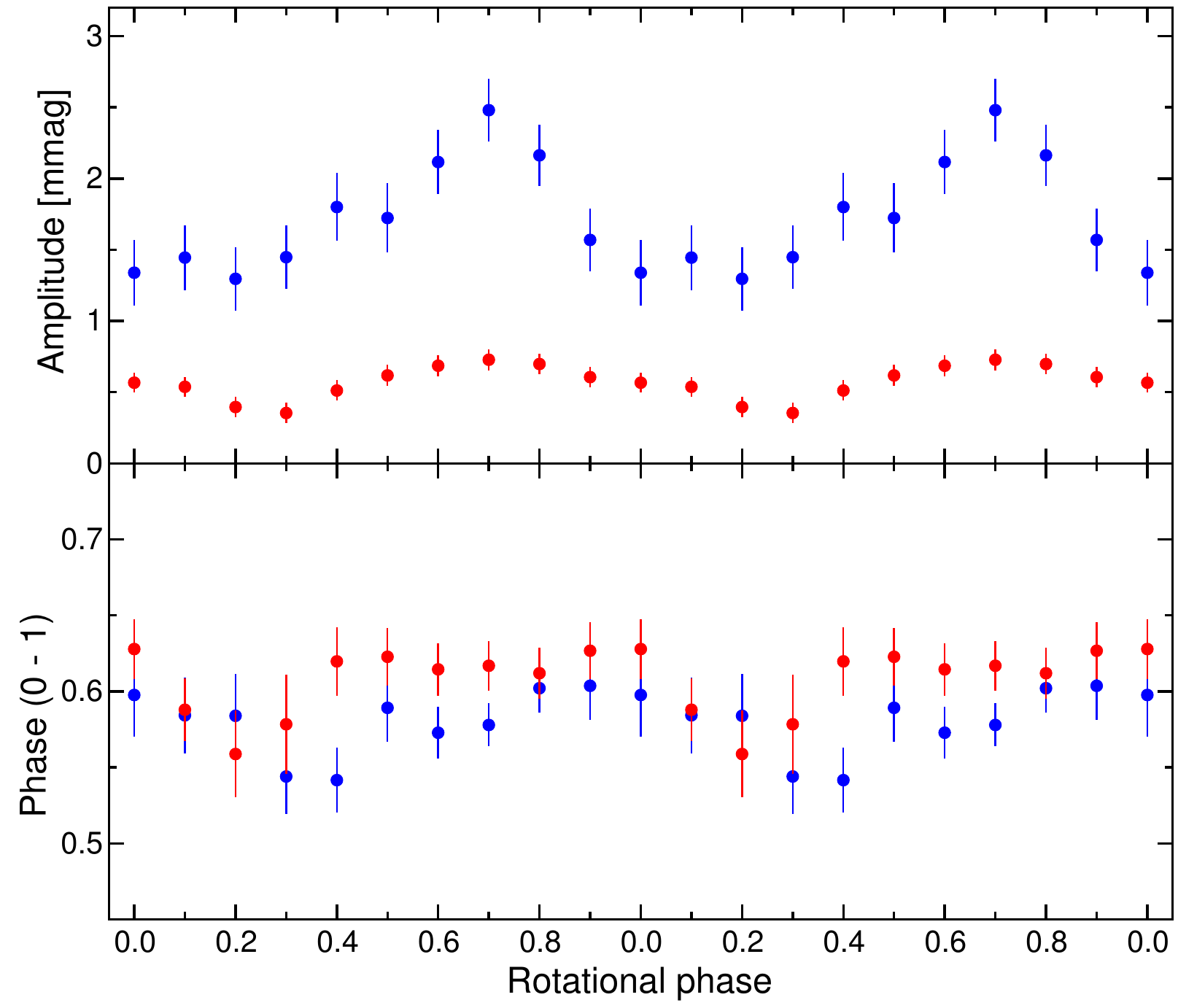}
     \caption{Change of amplitude and phase of $f_1$ with the rotation phase for the  BRITE-blue data (blue dots) and BRITE-red data (red dots) computed for the combined 2014 and 2016 data set.
                  }
     \label{f:ap3}
\end{figure}

With the new BRITE data, we  computed Fig.\,\ref{f:ap3}, which is  analogous to Fig.\,11 in Paper\,I, showing the amplitude and phase modulation of $f_1$ with the rotation phase. The phases given in this new figure differ from those in Paper\,I because different rotation periods have been used. The data are binned into ten rotation phases, where the starting epoch and the rotation period are the same as were used for phasing the rotational changes. The nearly sinusoidal shape of the amplitude modulation and the practically constant phase is consistent with an axisymmetric $l=1, m=0$ mode \citep{bruntt2}, with a rather small tilt between the pulsation and rotation axis \citep{bigot}.

\subsection{Additional pulsation frequencies ($f_6$, $f_7$, and rotational side lobes of $f_1$)}

After pre-whitening $f_1$ from the combined BRITE data we still find some signal left at frequencies where \citet{bruntt2} reported $f_6$ and $f_7$ in the WIRE data and which are unambiguously confirmed by TESS (Fig.\,\ref{f:ap1}). Even though the corresponding peaks barely exceed a S/N of three in the individual 2014 and 2016 BRITE data, they reach sufficient significance levels in the combined data (Table\,\ref{t:freq}). Given the larger amplitude, $f_7$ reaches a probability of 0.99 (and therefore very strong evidence of not being due to noise) in both data sets. On the other hand, $f_6$ has a smaller amplitude and therefore also a lower probability of 0.91 (strong evidence) in the Bb$^{\star}$  data and 0.77 (still substantial evidence) in the BTr data. We note that we would not consider the last peak as real on its own, but given its significance beyond doubt in all other time series we accept that it is real in the BTr data. In a more Bayesian sense, one could use the TESS frequency (and its uncertainty) as a  prior in the frequency analysis of the BTr data. This would significantly increase the peak's probability of  not being due to noise,  but this  approach  considerably exceeds our present computational resources. 

Even though $f_6$ and $f_7$ are easily detectable in the combined BRITE data, we find again an aliasing problem when determining their best-fit parameters. A closer look at Fig.\,\ref{f:ap2} shows that while the strongest peak in the BTr spectrum is consistent with the position of $f_7$ in the TESS data, the dominant peak in the Bb$^{\star}$  spectrum is shifted towards higher frequencies. However, the Bb$^{\star}$16 time series has several large gaps;  the longest is about 12.9\,d long, which causes aliases at about $\pm0.078$\,\cd\ (in addition to the $\pm0.0019$\,\cd\ aliases due to the large gap between the 2014 and 2016 data). If we now force our frequency analysis algorithm to fit the alias at $-\,0.078$\,\cd\ instead of the largest-amplitude peak, we get very good agreement between the BTr and Bb$^{\star}$ data. This  strategy  might appear arbitrary, but in fact it represents the Bayesian principle of  prior information and is therefore justified.

We find a similar situation for $f_6$. While the largest-amplitude peak in the BTr spectrum agrees well with the position of the peak in the TESS data, the dominant peak in the Bb$^{\star}$ spectrum appears to be a $+\,0.0019$\,\cd\ alias. A forced fit again aligns the frequencies we find in the blue and red BRITE data.

Finally, we  also find  some signal in the Bb$^{\star}$ data at the expected rotational side lobes of $f_1$. The pulsation amplitude of an obliquely pulsating non-radial mode changes with  rotation phase, as the aspect of the mode changes. This gives rise to frequency side lobes $f_1 \pm f_\mathrm{rot}$ that describe the amplitude modulation (and phase variation, if present), which are clearly visible in the WIRE and TESS data. While the signal at $f_{1}-$ is too weak to be distinguished from the noise (even though we know it has to be there), the peak at $f_{1}+$ is with $p=0.82$ statistically significant. This allows for a seismic determination of the  \ac\  rotation period, which results in $4.4758\pm 0.0020$\,d, and is more accurate than those resulting from the WIRE and TESS data. Despite the greater noise of the BRITE data, the much longer time base permits  a more accurate determination of \prot.

\subsection{Mode identification and the large frequency separation}

A frequently used observable in asteroseismic studies of high-overtone acoustic oscillations is the large frequency separation  $\Delta\nu_{n,l}$, which is defined as the difference between modes of the same spherical degree and consecutive radial orders: $\nu_{n+1,l} - \nu_{n,l}$. The large separation becomes relevant for high radial orders, which are expected to follow the asymptotic relation \citep{tas1980}: 
\begin{equation} \label{eq:dnu}
\nu_{n,l} \simeq \Delta\nu _{n,l}\, (n + l/2 + \epsilon_0) - l(l+1)D_0. 
\end{equation}
Here $\epsilon_0$ and $D_0$ are parameters sensitive to the properties of the reflection layer near the stellar surface and the conditions in the stellar core, respectively. But more importantly, the large separation is related to the stellar acoustic diameter (i.e. inverse sound travel time across the stellar diameter). For an ideal adiabatic gas, this is proportional to the square root of the mean stellar density. Consequently, $\Delta\nu$ provides a measure for the mean density of a star.  

Even though a strong magnetic field has the tendency to distort spherical symmetry, the high-frequency oscillations observed in \ac\ indicate that these modes are high-overtone acoustic oscillations for which one may expect to find a regular pattern corresponding to the asymptotic relation. As already noted by \citet{bruntt2}, the three modes $f_1$, $f_6$, and $f_7$ are almost equidistant in frequency with an average separation of $\sim\,$2.606\,\cd . One can naively expect that this value corresponds to the average frequency separation of \ac . However, it is not compatible with other constraints on the stellar properties as we show in the following. 

\citet{bruntt1} determined the angular diameter of \ac\ to $\Theta_{LD}\,=\,1.105\pm0.037$\,mas, and used the revised Hipparcos parallax of \citet{vanLeu2007} to estimate the star's radius to be $1.97\pm0.07$\,R$_\sun$. 
With an average frequency separation of $\sim\,$2.606\,\cd\  one can estimate the star's mass according to $\Delta\nu \propto \sqrt{M/R^3}$.
This is usually done by relating $\Delta\nu$ to the solar value of about 11.664\,\cd\ \citep[e.g.][]{kal2010}, which gives an unrealistically low mass of $0.38\pm0.04$\,M$_\sun$. However, if we follow \citet{bruntt2} and set $\Delta\nu = 5.21$\,\cd\  (i.e. $f_7$ - $f_6$), we obtain a plausible mass of $1.52\pm0.15$\,M$_\sun$, which is also compatible with the estimate of $1.7\pm0.2$\,M$_\sun$, based on the star's position in the HR diagram \citep{bruntt1}. 
We expect \ac\ to have a slightly different mass than 1.5\,M$_\sun$ because the observed modes are magnetically distorted \cite[e.g.][]{cun2006}, which also affects the large separation compared to that of unperturbed modes for which the $\Delta\nu$ scaling is defined.
 
\begin{figure}[tbp]
     \includegraphics[width=\hsize]{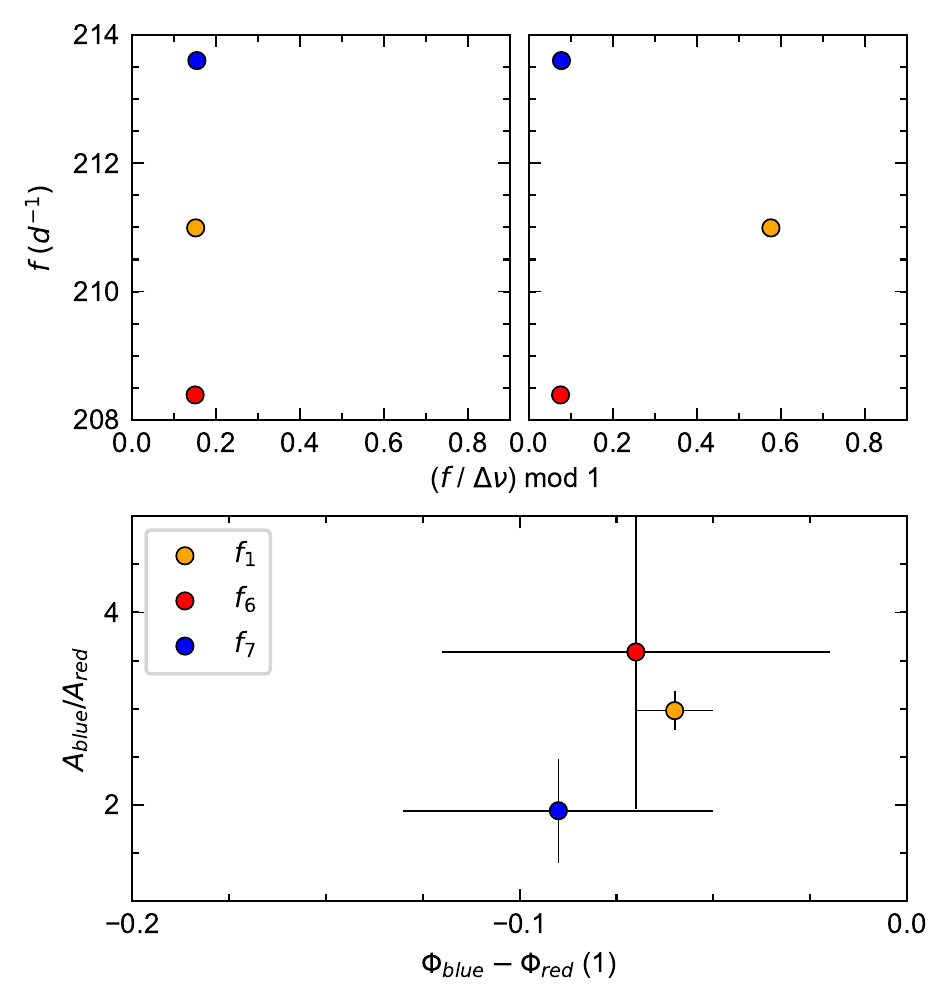}
     \caption{\textit{Top}: Echelle diagrams including the three frequencies detected in the BRITE data with a large separation $\Delta\nu = 2.6$\,\cd (left), and 5.2\,\cd (right). \textit{Bottom}: Amplitude ratios vs. phase differences in the two BRITE bands for the same frequencies.
                  }
     \label{f:ap4}
\end{figure}

A consequence of  $\Delta\nu$ being more likely equal to 5.21\,\cd\ than to 2.606\,\cd\ is that the three modes discussed above cannot be of the same spherical degree. This is also supported by Fig.\,\ref{f:ap4}, where we show the three modes in an Echelle diagram. While the modes almost perfectly line up vertically when folding their frequency with 2.6\,\cd\ (indicating the same spherical degree), this is not so when folding them with 5.2\,\cd . In this case, $f_1$ is shifted by slightly more than one-half in the horizontal direction, while the other two modes have about the same horizontal offset. This is  expected for a sequence of two consecutive $l=0$ modes with an intermediate $l=1$ mode, or vice versa (Eq.\,\ref{eq:dnu}). Assuming that the magnetic distortion is similar for all three modes, this also rules out the presence of a quadruple mode. Such a mode would need to be located closer to one mode than to the other (and not at their midpoint). Since \citet{bruntt2} argued that $f_1$ is very likely an axisymmetric dipole mode ($l=1, m=0$) based on simulated amplitude modulations for an oblique pulsator model, we conclude that $f_6$ and $f_7$ are very likely consecutive radial modes.

Even though a mode identification for roAp stars from multicolour photometry is notoriously difficult \citep[e.g.][]{Quitral}, further support is provided by the amplitude ratios and phase differences in the two BRITE passbands we determine for the three modes (Fig.\,\ref{f:ap4}). We   expect that modes of the same spherical degree form clusters in a diagram like  Figure\,\ref{f:ap4}. If we find modes that are clearly separated in the amplitude-ratio versus phase-difference plane, they can be assumed to have different spherical degrees. For $f_6$ the uncertainties are too large to make any conclusions, but $f_1$ and $f_7$ are separated by about $1.8\,\sigma$ in their amplitude ratio, which corresponds to a probability of almost 0.9 that they are  separated, and therefore  have a different spherical degree.

\section{Discussion}  

-- {\sc Data:}
We analyse and discuss data obtained by \bc\ during 2014 and 2016, covering a total of 316 days and complement data volume and time base with observations archived from \w , \s , and \t .
These data suffer from different noise levels and instrumental effects, which need to be considered and corrected. Corrections of trends and averaging procedures need to be optimised for a discussion of long-term (rotation, spots) and short-term (pulsation) effects. An attractive aspect of the \b\ data is the availability of data in two passbands, some of them taken nearly simultaneously. This allowed us to study in detail the pros and cons of Bayesian based photometric surface imaging.

Not surprisingly, the photometric quality of the data seriously limits the number of detectable spots and their location, as is discussed in Section \ref{s:spot}. Spot longitudes are well constrained by the data; instead, latitudes are, photometrically, a notoriously ill-defined parameter. In the case of Doppler imaging, the situation is much better due to the availability of additional information. 

-- {\sc Rotation:}
The determination of \prot\ for \ac\ dates back to 1991 \citep{mathys2}. Meanwhile, many photometric, spectroscopic, and polarimetric observation were performed and we determined the formally most accurate \prot\ from combined \w\ and \t\  spot transit times to be 4.47930\,$\pm 0.00002$\,d  (68\%\ interval).  

The different shape of the rotation-phase plot extracted from the red and blue data indicates a different chemical composition of at least one of the spots, which, however, can only be tested by high-resolution spectroscopy.

-- {\sc Spots:}
Two spots were identified in Paper\,I in the red \b\ data from 2014, but with better data obtained with BTr in 2016 we have clear evidence of three spots. Despite three more degrees of freedom, the gain in evidence from a two-spot model to a three-spot model is substantial, although the residuals to the model light curves decreased only from 1.389 to 1.386 mmag. In comparison, the much noisier data from \s\ (up to a factor of 4) allow only two spots to be considered.

Bayesian photometric imaging routines result in  numerous solutions. As an example,  Fig.\,\ref{f:allmaps} shows two solutions to the BTr (2016) data: the barycentric solution (top, mean values) and best-fit solution (bottom, modal values). They do not coincide satisfactorily, which hints at a serious non-Gaussianity of the posterior, as is illustrated in Fig.\,\ref{f:mcmc}. The most striking difference between the Bayesian photometric imaging of the BTr (2016) photometry
and the WIRE data set (see Fig.\,\ref{f:allmaps}) concerns spot latitudes. The marginal distribution of the spot\,1 latitude for the BTr data seems to exclude  a northern location, contrary to what is determined for the WIRE data. The \lq north pole\rq\ is defined as the pole that  is visible from Earth.

Here a cautionary note should be heeded: 
One should keep in mind that the often surprisingly narrow marginal distribution is due to a
model's rigidity. It measures its \lq elbow room\rq\ and  nothing else. If one could relax the rigidity of the model, for example  by allowing non-circular spots and/or variable contrasts, the marginal distribution would spread out because of the increased freedom.
But any additional degrees of freedom come at a price: they would reduce the  evidence of a three-spot model. There is a trade-off between precision of a fit and its credibility. We adhere here to Fermi's rule: it is better to be approximately right than precisely wrong.

-- {\sc Pulsation:}
No pulsation is detected in the SMEI data with an amplitude exceeding the detection threshold of 0.33\,mmag, corresponding to a S/N = 4.5. The dominant pulsation frequency $f_1$ derived from  BRITE-red data agrees exceptionally well with that derived from \b-blue data, but there are significant differences to the pulsation frequencies derived from \w\ and \t , which will   be subject to a follow-up paper with a detailed asteroseismic analysis (Kallinger et al., in preparation).
We can improve the pulsation frequency by combining the times of maximum from BRITE-red and WIRE data, resulting in $f_1$\,=\,210.993264(5)\,d$^{-1}$, which  is the most accurately determined $f_1$ to date for any roAp star.
The three hitherto well-established frequencies very likely come from two consecutive radial $l\,=\,0$ modes ($f_6\,\rm{and}\ f_7$), with an intermediate $l\,=\,1$ mode ($f_1)$.

\section{Conclusion}
At least three surface spots can be identified for \ac , which confirms  the conclusion of Paper\,I that a two-spot model  is too simple. The high-quality \t\ data even suggest  a fourth spot.  On the other hand, the low-quality blue \b\  data barely indicate a spot at rotation phase 0.6 (Fig.\,\ref{f:rot}). 

According to our experience the  best-fit set of parameters, indicated by a minimum $\chi ^2$, differs significantly from the set of mean values (inferred from the marginal distributions of the parameters), which hints at a noticeable skewness of the probability distribution in the ten-dimensional configuration space considered. 
Spot latitudes are,  as expected,  less well determined than longitudes.
To our knowledge, this is the first time 
that a Bayesian-based evidence of models differing in the number of spots has been quantitatively determined. 

Concerning the main pulsation frequency  of \ac , we were able to improve  the accuracy to 60\,pHz (0.01\,msec for a 6.825\,min period), assuming a stable frequency. 

The photometric data obtained to date  for \ac\ clearly illustrate   the need for high-precision data on the one hand {and} long data sets on the other. In general, both qualities are needed for convincing astrophysical analyses, and this  should  be considered when deciding about investments in ground-based and  in space-based instrumentation.

\begin{acknowledgements}
The authors are grateful to 
Andrzej Pigulski, Hans Bruntt, Margarida Cunha, and Denis Shulyak for providing valuable input to this investigation. We also thank an anonymous referee for a careful report, which helped to improve the paper.
Adam Popowicz was supported by Silesian University of Technology, Rector Grant 02/140/RGJ20/0001 (image processing and automation of photometric routines of BRITE-nanosatellite data). GH acknowledges financial support by the Polish NCN grant 2015/18/A/ST9/00578. 
AFJM and GW are grateful to NSERC (Canada) for financial aid.
The MCMC computations have been performed by HEF at the AIP. 
This paper includes data collected by the TESS mission, funded by the NASA Explorer Program.
\end{acknowledgements}

\bibliographystyle{aa}
\bibliography{38345.bib}

\end{document}